\documentclass[preprint,nofootinbib,tightenlines,groupedaddress,amsmath,amssymb]{revtex4}

\usepackage{bm}% bold math
\usepackage{color}
\usepackage{graphicx}% Include figure files
\usepackage[hypertex]{hyperref}
\begin{document}

\vspace*{3em}

\title{Large electron electric dipole moment in minimal flavor violation
framework with Majorana neutrinos}
\author{Xiao-Gang He,$^{1,2,3}$ Chao-Jung Lee,$^{3}$ Siao-Fong Li,$^{3}$ and Jusak Tandean$^{3}$}
\affiliation{${}^{1}$INPAC, SKLPPC, and Department of Physics,
Shanghai Jiao Tong University, Shanghai 200240, China \vspace{3pt} \\
${}^{2}$Physics Division, National Center for
Theoretical Sciences, Department of Physics, National Tsing Hua
University, Hsinchu 300, Taiwan \vspace{3pt} \\
${}^{3}$CTS, CASTS, and
Department of Physics, National Taiwan University, Taipei 106, Taiwan \vspace{3ex}}

%\date{\today $\vphantom{\bigg|_{\bigg|}^|}$}

\begin{abstract}
The latest data from the ACME Collaboration have put a stringent constraint on the electric
dipole moment $d_e$ of the electron.
Nevertheless, the standard model (SM) prediction for $d_e$ is many orders of magnitude below
the new result, making this observable a powerful probe for physics beyond the SM.
We carry out a~model-independent study of $d_e$ in the SM with right handed neutrinos and its
extension with the neutrino seesaw mechanism under the framework of minimal flavor violation.
We find that $d_e$ crucially depends on whether neutrinos are Dirac or Majorana particles.
In the Majorana case, $d_e$ can reach its experimental bound, and it constrains the scale of
minimal flavor violation to be above a few hundred GeV or more.
We also explore extra $CP$-violating sources in the Yukawa couplings of the right-handed neutrinos.
Such new sources can have important effects on $d_e$.
\end{abstract}

\maketitle

Understanding the origin of $CP$ violation is one of the outstanding challenges of modern
particle physics.
It may hold the key to solving the problem of why our Universe is dominated by matter over
antimatter.
Although $CP$ violation has been observed in flavor-violating processes such as the mixing
of neutral $K$ and $B$ mesons with their respective antiparticles and the decays of these
particles~\cite{pdg}, no laboratory experiments have yet revealed evidence of $CP$ violation
in flavor-conserving transitions~\cite{He:1989xj,suzuki,acme}.
The best constraints for the latter class of processes are available from measurements on
the electric dipole moments (EDMs) of fermions, such as the electron and neutron.
A~nonzero EDM breaks time-reversal $(T)$ and $CP$ symmetries.
Since the standard model (SM) predictions for electron and neutron EDMs are way below their
current experimental bounds~\cite{suzuki,acme,hesm,Pospelov:2013sca}, EDM searches are
powerful probes for new sources of $T$ and $CP$ violation beyond the SM.

Recently the ACME Collaboration~\cite{acme}, looking for the electron EDM $d_e$ using the polar
molecule thorium monoxide, reported a~new result of
\,$d_e=(-2.1\pm3.7_{\rm stat} \pm 2.5_{\rm syst})\times 10^{-29\;}e$\,cm,\, corresponding to
an upper bound of \,$|d_e|< 8.7\times 10^{-29\,}e$\,cm\, at the 90\% confidence level.
This is an improvement over the previous strictest bound by an order of magnitude, but still
much higher than the SM expectation for~$d_e$, of order $10^{-44\,}e$\,cm~\cite{Pospelov:2013sca}.

Beyond the SM, $d_e$ can greatly exceed its SM prediction and even reach its measured
bound~\cite{suzuki}.
Such enhancement may hail from various origins depending on the specifics of the new physics models.
Therefore, it is desirable to do an analysis of $d_e$ which treats some general features of
the new physics without going into detailed model structures.
In the literature, there is indeed a~framework, that of minimal flavor
violation (MFV)~\cite{D'Ambrosio:2002ex,Cirigliano:2005ck}, where such an analysis can be performed.
Under the assumption of~MFV, the sources of all flavor-changing neutral currents (FCNC) and $CP$
violation reside in renormalizable Yukawa couplings defined at the tree level.

In this paper, we adopt the MFV hypothesis to examine $d_e$ in the SM with right-handed neutrinos
and its extension with neutrino seesaw mechanism.
We look at MFV contributions to $d_e$ via effective dipole interactions.
We find that its size depends considerably on whether neutrinos are Dirac or Majorana fermions.
In the Majorana case, $d_e$ can reach its experimental bound, which thus limits the MFV scale to
above a few hundred GeV or more.

In the SM that is slightly expanded with the addition of three right-handed neutrinos,
the renormalizable Lagrangian relevant to lepton masses is
\begin{eqnarray} \label{Lm}
{\cal L}_{\rm m}^{} \,\,=\,\,  - \bar L_{i,L\,}^{}(Y_\nu)_{ij\,}^{}\nu_{j,R\,}^{}\tilde H
\,-\, \bar L_{i,L\,}^{}(Y_e)_{ij\,}^{}E_{j,R\,}^{}H  \,-\, \mbox{$\frac{1}{2}$}\,
\overline{\nu^{\rm c}}_{\!\!\!i,R}^{}\,(M_\nu)_{ij\,}^{}\nu_{j,R}^{}
\;+\; {\rm H.c.} ~,
\end{eqnarray}
where \,$i,j=1,2,3$\, are summed over, $L_{i,L}$ denotes left-handed lepton doublets,
$\nu_{j,R}^{}$ $(E_{j,R})$ represents right-handed neutrinos (charged leptons),
$Y_{\nu,e}$ are matrices for the Yukawa couplings, $H$ is the Higgs doublet,
\,$\tilde H=i\tau_2^{}H^*$,\, and $M_\nu$~is the Majorana mass matrix for~$\nu_{j,R}^{}$.
The $M_\nu$ part facilitates the seesaw mechanism for light neutrino masses~\cite{seesaw},
but it is absent if neutrinos are Dirac particles.

The basics of the MFV framework are given in\,\,\cite{D'Ambrosio:2002ex,Cirigliano:2005ck}.
For the lepton sector, if neutrinos are Dirac fermions, the MFV hypothesis implies that
${\cal L}_{\rm m}$ in Eq.\,(\ref{Lm}) is formally invariant under the~global
group~\cite{Cirigliano:2005ck}
\,${\rm U}(3)_L\times{\rm U}(3)_\nu\times{\rm U}(3)_E =
G_\ell\times{\rm U}(1)_L\times{\rm U}(1)_\nu\times{\rm U}(1)_E$,\,
with \,$G_\ell ={\rm SU}(3)_L\times{\rm SU}(3)_\nu\times{\rm SU}(3)_E$,\, while $L_{i,L}$,
$\nu_{i,R}$, and $E_{i,R}$ transform as fundamental representations of SU$(3)_{L,\nu,E}$,
respectively.
In the Majorana case, the $M_\nu$ term in Eq.\,(\ref{Lm}) changes the global group in that
the U(3)$_\nu$ is broken into O(3)$_\nu$ if $\nu_{i,R}$ are degenerate 
or that the U(3)$_\nu$ is completely broken if $\nu_{i,R}$ are nondegenerate.

We will work in the basis where $Y_e$ is already diagonal, namely,
\,$Y_e=\sqrt2\, {\rm diag}\bigl(m_e^{},m_\mu^{},m_\tau^{}\bigr)/v$,\, with \,$v\simeq246$\,GeV\,
being the vacuum expectation value of $H$, and $\nu_{L,R}$ and $E_{L,R}$ refer to the mass
eigenstates.  If neutrinos are of Dirac nature, we can express $L_{i,L}$ and $Y_\nu$ in terms of
the Pontecorvo-Maki-Nakagawa-Sakata mixing matrix $U_{\scriptscriptstyle\rm PMNS}$ as
\begin{eqnarray} \label{Ynu}
L_{i,L}^{} \,= \left( \!\begin{array}{c} (U_{\scriptscriptstyle\rm PMNS})_{ij\,}^{} \nu_{j,L}^{}
\vspace{2pt} \\
E_{i,L}^{} \end{array}\! \right) , ~~~~~~~
Y_\nu \,=\, \frac{\sqrt2}{v}\,U_{\scriptscriptstyle\rm PMNS}^{}\,\hat m_\nu^{} ~,
\end{eqnarray}
where \,$\hat m_\nu^{}={\rm diag}\bigl(m_1^{},m_2^{},m_3^{}\bigr)$\, involves the light neutrino
eigenmasses $m_{1,2,3}^{}$.

If neutrinos are Majorana fermions, $Y_\nu$ needs to be modified.
The presence of \,$M_\nu\gg v Y_\nu/\sqrt{2}$\, in Eq.\,(\ref{Lm}) activates the seesaw
mechanism~\cite{seesaw}, leading to the light neutrinos' mass matrix
\begin{eqnarray} \label{ligh-neu}
m_\nu \,\,=\,\, -\frac{v^2}{2}\, Y_\nu^{}M_\nu^{-1}Y_\nu^{\rm T} \,\,=\,\,
U_{\scriptscriptstyle\rm PMNS\,}^{}\hat m_{\nu\,}^{}U_{\scriptscriptstyle\rm PMNS}^{\rm T} ~.
\end{eqnarray}
This allows one to choose $Y_\nu$ to be~\cite{Casas:2001sr}
\begin{eqnarray} \label{ym}
Y_\nu^{} \,\,=\,\,
\frac{i\sqrt2}{v}\,U_{\scriptscriptstyle\rm PMNS\,}^{}\hat m^{1/2}_\nu OM_\nu^{1/2}~,
\end{eqnarray}
where $O$ satisfies \,$OO^{\rm T}=\openone$,\, which is a 3$\times$3 unit matrix, and
\,$M_\nu={\rm diag}(M_1,M_2,M_3)$.\,

To derive nontrivial two-lepton FCNC and $CP$-violating interactions, one assembles an arbitrary
number of the Yukawa matrices to devise the $G_\ell$ representations
\,$\Delta_\ell \sim (1\oplus8,1,1)$, \,$\Delta_{\nu 8}\sim (1,1\oplus8,1)$,
\,$\Delta_{e8}\sim (1,1,1\oplus8)$, \,$\Delta_\nu \sim (\bar 3, 3, 1)$,\, and
\,$\Delta_e\sim(\bar 3, 1, 3)$,\, combines them with lepton fields to arrive at
the $G_\ell$-invariant objects \,$\bar L_L\Delta_\ell L_L$,\, $\bar \nu_R \Delta_{\nu 8} \nu_R$,
\,$\bar E_R \Delta_{e8}E_R$,\, $\bar \nu_R \Delta_\nu L_L$,\, and~\,$\bar E_R \Delta_e L_L$,\,
and attaches appropriate numbers of the Higgs and gauge fields to form singlets under the SM
gauge group.
Since fermion EDMs flip chirality, only the last combination, $\bar E_R \Delta_e L_L$, is of
interest to our examination of $d_e$. We can write \,$\Delta_e=Y^\dagger_e\Delta$,\, with $\Delta$
built up of terms in powers of \,${\sf A}=Y_\nu^{}Y_\nu^\dagger$\,
and~\,${\sf B}=Y_e^{}Y^\dagger_e$.\,

Formally, $\Delta$ consists of an infinite number of terms,
\,$\Delta=\sum\xi_{ijk\cdots}^{}\,{\sf A}^i{\sf B}^j{\sf A}^k\cdots$.\,
Under the MFV assumption, the coefficients $\xi_{ijk\cdots}^{}$ must be real because complex 
$\xi_{ijk\cdots}$ would introduce new $CP$-violating sources beyond what is already contained 
in $Y_{e,\nu}$.
Using the Cayley-Hamilton identity
\,$X^3=X^2\,{\rm Tr}X+X\bigl[{\rm Tr}X^2-({\rm Tr}X)^2\bigr]/2+\openone{\rm Det}X$\, for
a~$3\times 3$ matrix~$X$, this infinite series can be resummed into a~finite number of
terms~\cite{Colangelo:2008qp}:
\begin{eqnarray} \label{general}
\Delta &\,=\,& \xi^{}_1\openone + \xi^{}_{2\,}{\sf A}+\xi^{}_{3\,}{\sf B}
+ \xi^{}_{4\,}{\sf A}^2 +\xi^{}_{5\,}{\sf B}^2 + \xi^{}_{6\,}{\sf AB} + \xi^{}_{7\,}{\sf BA}
+ \xi^{}_{8\,}{\sf ABA} + \xi^{}_{9\,}{\sf BA}^2 + \xi^{}_{10\,}{\sf BAB}
\nonumber \\ && \! +~
\xi^{}_{11\,}{\sf AB}^2 + \xi^{}_{12\,}{\sf ABA}^2 + \xi^{}_{13\,}{\sf A}^2{\sf B}^2
+ \xi^{}_{14\,}{\sf B}^2{\sf A}^2 + \xi^{}_{15\,}{\sf B}^2{\sf AB}
+ \xi^{}_{16\,}{\sf AB}^2{\sf A}^2+\xi^{}_{17\,}{\sf B}^2{\sf A}^2{\sf B} ~. ~~~
\end{eqnarray}
Although $\xi_{ijk\cdots}^{}$ are real, the reduction of the infinite series into the 17 terms
can make the coefficients $\xi_r^{}$ in Eq.\,(\ref{general}) complex due to imaginary parts
among the traces of the matrix products~\,${\sf A}^i{\sf B}^j{\sf A}^k\cdots$.\,
Such imaginary contributions are proportional to the Jarlskog invariant
\,${\rm Im\,Tr}\bigl({\sf A}^2{\sf BAB}^2\bigr) =
(i/2)\,{\rm Det}[{\sf A,B}]\ll1$\,~\cite{Colangelo:2008qp}.
This implies that, with all $\xi_{ijk\cdots}^{}$ being of ${\cal O}$(1), the contributions of
$\xi_r^{}$ to $d_e$ are suppressed by a factor of \,$m^2_\mu m^2_\tau/v^4$\, compared to
the contribution from ${\sf ABA}^2$ which has the least number of the suppressive factor $Y_e$
among the products in Eq.\,(\ref{general}) that contribute to~$d_e$.
Therefore, hereafter we neglect~${\rm Im}_{\,}\xi_r^{}$.

The EDM $d_l$ of a lepton $l$ is described by
\,${\cal L}_d=-(i d_l/2)\bar l \sigma^{\kappa\omega}\gamma_5 l F_{\kappa\omega}$,\,
where $F_{\kappa\omega}$ is the photon field strength tensor.
In the MFV framework, this arises at lowest order from the operators in the effective
Lagrangian~\cite{Cirigliano:2005ck}
\begin{eqnarray} \label{Leff}
{\cal L}_{\rm eff} \,\,=\,\, \frac{1}{\Lambda^2} \Bigl( g'\bar E_R^{}Y^\dagger_e
\Delta_{1\,}^{}\sigma_{\kappa\omega}^{}H^\dagger L_L^{}B^{\kappa\omega}
\,+\, g\bar E_R^{}Y^\dagger_e\Delta_{2\,}^{}\sigma_{\kappa\omega}^{}H^\dagger\tau_j^{}L_L^{}
W_j^{\kappa\omega} \Bigr) \;+\; {\rm H.c}. ~,
\end{eqnarray}
where $\Lambda$ is the MFV scale, $W$ and $B$ denote the SM gauge fields with coupling
constants $g$ and~$g'$, respectively, $\tau_j^{}$~are Pauli matrices, and $\Delta_{1,2}$ are
of the form in~Eq.\,(\ref{general}) with generally different~$\xi_r^{(1,2)}$.

Expanding Eq.\,(\ref{Leff}), one can identify the contributions to the charged lepton EDMs.
They are proportional to \,Im$\bigl(Y^\dagger_e\Delta_{1,2}\bigr){}_{kk}^{}$,\, but we find
that in
$\Delta_{1,2}$ only the terms proportional to  ${\sf ABA}^2$ and ${\sf AB}^2{\sf A}^2$
are pertinent.  Thus, for the electron
\begin{eqnarray} \label{de}
d_e \,=\, \frac{\sqrt2\,e_{\,}v}{\Lambda^2} \Bigl[
\xi^\ell_{12\,}{\rm Im}\bigl(Y^\dagger_e {\sf ABA}^2\bigr)_{11}
+ \xi^\ell_{16\,} {\rm Im}\bigl(Y^\dagger_e {\sf AB}^2{\sf A}^2\bigr)_{11} \Bigr]  ~,
\end{eqnarray}
where  \,$\xi^\ell_r=\xi^{(1)}_r-\xi^{(2)}_r$.\,
Accordingly, with only one generation of leptons, $d_e$ would be identically zero
whether neutrinos are Dirac or Majorana particles.
If neutrinos are Dirac particles, we obtain from Eqs.\,\,(\ref{Ynu}) and (\ref{de})
\begin{eqnarray} \label{dedirac}
d_e^{\rm D} \,=\, \frac{32 e_{\,}m_e^{}}{\Lambda^2} \Biggl[ \xi^\ell_{12} +
\frac{2\bigl(m_\mu^2+m^2_\tau\bigr)}{v^2}\,\xi^\ell_{16} \Biggr] 
\frac{\bigl(m_\mu^2-m_\tau^2\bigr)\bigl(m^2_1-m_2^2\bigr)\bigl(m_2^2-m_3^2\bigr)
\bigl(m_3^2-m_1^2\bigr)}{v^8}\,J_\ell^{} \,,
\end{eqnarray}
where \,$J_\ell={\rm Im}\bigl(U_{e2}^{}U_{e3}^*U^*_{\mu2}U_{\mu 3}^{}\bigr)$\, is
the Jarlskog invariant for $U_{\scriptscriptstyle\rm PMNS}$.
In the Majorana neutrino case, if $\nu_{i,R}$ are degenerate, \,$M_\nu={\cal M}\openone$,\,
and $O$ is a~real orthogonal matrix, from Eq.\,(\ref{ym}) we have
\,${\sf A}=(2/v^2){\cal M}_{\,}U_{\scriptscriptstyle\rm PMNS\,}^{}\hat m^{}_\nu
U_{\scriptscriptstyle\rm PMNS}^\dagger$\,
and consequently
\begin{eqnarray} \label{dem}
d_e^{\rm M} \,\,=\,\, \frac{32e_{\,}{\cal M}^{3\,}m_e^{}}{\Lambda^{2\,}v^8}
\bigl(m_\mu^2-m_\tau^2\bigr) \bigl(m_1^{}-m_2^{}\bigr)\bigl(m_2^{}-m_3^{}\bigr)
\bigl(m_3^{}-m_1^{}\bigr)\, \xi_{12\,}^\ell J_\ell^{} ~,
\end{eqnarray}
neglecting the $\xi^\ell_{16}$ term.  Clearly $d_e^{\rm M}$ is substantially enhanced
relative to $d_e^{\rm D}$ due to ${\cal M}m_i^{}\gg m^2_i$.

In the preceding cases, $d_e$ comes from the $CP$-violating Dirac phase $\delta$ in
$U_{\scriptscriptstyle\rm PMNS}$, and the two Majorana phases $\alpha_{1,2}$ therein
do not enter.
However, if $\nu_{i,R}$ are not degenerate, nonzero $\alpha_{1,2}$ can lead to an extra effect
on $d_e$ even with a real \,$O\neq\openone$.\,
With a complex $O$, its phases may induce an additional contribution to $d_e$,
whether or not $\nu_{i,R}$ are degenerate.
We will explore these scenarios numerically later.

To evaluate $d_e$, we need to know the $U_{\scriptscriptstyle\rm PMNS}$ elements.
In the standard parametrization of\,\,\cite{pdg}, it depends on three mixing angles
$\theta_{12,13,23}$ and the phase $\delta$ for Dirac neutrinos.
We adopt their values from the latest fit to the global data on neutrinos with a normal
hierarchy (NH) or inverted hierarchy (IH) of masses in\,\,\cite{Capozzi:2013csa},
which also provides their square differences.
For Majorana neutrinos, $U_{\scriptscriptstyle\rm PMNS}$ includes a~phase matrix 
\,$P={\rm diag}\bigl(e^{i\alpha_1/2},e^{i\alpha_2/2},1\bigr)$\, multiplied from 
the right~\cite{pdg}.  Since $\alpha_{1,2}$ are still unknown, we will select specific values 
for them in our illustrations.

For Dirac neutrinos, we scan the empirical ranges of the parameters
from\,\,\cite{Capozzi:2013csa} to maximize $d_e^{\rm D}$ in~Eq.\,(\ref{dedirac}).
The result in the NH or IH case is
\,$d_e^{\rm D}=1.3\times10^{-99}\,\xi_{12}^\ell\,\bigl({\rm GeV}^2/\Lambda^2\bigr)\,e$\,cm.\,
This is negligible relative to the latest experimental upper bound~\cite{acme}.

If neutrinos are Majorana fermions, we now demonstrate that, in contrast, $d_e$ can be sizable.
We start with the simplest possibility that $\nu_{i,R}$ are degenerate,
\,$M_\nu={\cal M}\openone$,\, and the $O$ matrix in Eq.\,(\ref{ym}) is real.
Thus $d_e$ is already given in Eq.\,(\ref{dem}).
Scanning again the empirical parameter ranges in\,\,\cite{Capozzi:2013csa} to maximize
$d_e^{\rm M}$, we then obtain for \,$m_1^{}=0$\, $\bigl(m_3^{}=0\bigr)$ in the NH (IH) case
\begin{eqnarray} \label{deM'}
\!\!\frac{d_e^{\rm M}}{e\,\rm cm} \,\,=\,\, 47\;(5.2)\times10^{-24}\biggl(
\frac{{\cal M}}{10^{15\,}\rm GeV}\biggr)^{\!3}\biggl(\frac{\rm GeV}{\hat\Lambda}\biggr)^{\!2} ,
\end{eqnarray}
where \,$\hat\Lambda=\Lambda/|\xi_{12}^\ell|^{1/2}$\, and ${\cal M}$ is specified below.
Hence \,$|d_e^{\rm exp}|<8.7\times10^{-29}\;e$\,cm~\cite{acme} implies
\begin{eqnarray} \label{lambda}
\hat\Lambda \,\,>\,\,
0.74\;(0.24){\rm\;TeV}\;\biggl(\frac{{\cal M}}{10^{15\,}\rm GeV}\biggr)^{\!3/2} .
\end{eqnarray}

Since $d_e$ in Eq.\,(\ref{deM'}) is proportional to ${\cal M}^3$, one might naively think that
$d_e$ can easily reach its measured bound, which would therefore constrain $\hat\Lambda$ to
a~very high level with a very large ${\cal M}$.
However, there are restrictions on ${\cal M}$.
One is from consideration of the convergence of the series in Eq.(\ref{general}) which
supposedly includes arbitrarily high powers of $\sf A$ and $\sf B$.
If the biggest eigenvalue of $\sf A$ exceeds~1, the coefficients $\xi_r^{}$ might not converge
to finite numbers.
However, the expansion quantities may not necessarily be $\sf A$ and $\sf B$, depending on
the origin of~MFV.
It may be that it emerges from calculations of SM loops, in which case the expansion quantities
may more naturally be ${\sf A}/(16\pi^2)$ and ${\sf B}/(16\pi^2)$, restricting the eigenvalues
of $\sf A$ to below $16\pi^2$.
One may also consider instead the perturbativity condition for the Yukawa couplings, 
\,$(Y_\nu)_{ij}<\sqrt{4\pi}$\,~\cite{Kanemura:1999xf},
implying a cap on the eigenvalues of $\sf A$ at $4\pi$.
If we take the possible requirement that the eigenvalues of $\sf A$ not exceed~1 or~$4\pi$,
then \,${\cal M}=6.2\times10^{14}$\,GeV\, or \,$7.7\times10^{15}$\,GeV,\, respectively,
in~Eq.\,(\ref{deM'}), which are roughly the expected seesaw scales in some grand unified
theories.  Hence, \,$\hat\Lambda>360\,(120)\,$GeV\, or~\,$16\,(5)\,$TeV,\, respectively,
from Eq.\,(\ref{lambda}).
These $\cal M$ and $\hat\Lambda$ numbers would decrease if \,$m_{1(3)}>0$.\,
Hereafter, we require the eigenvalues of $\sf A$ to be less than~1.

\begin{figure}[b]
\includegraphics[width=87mm]{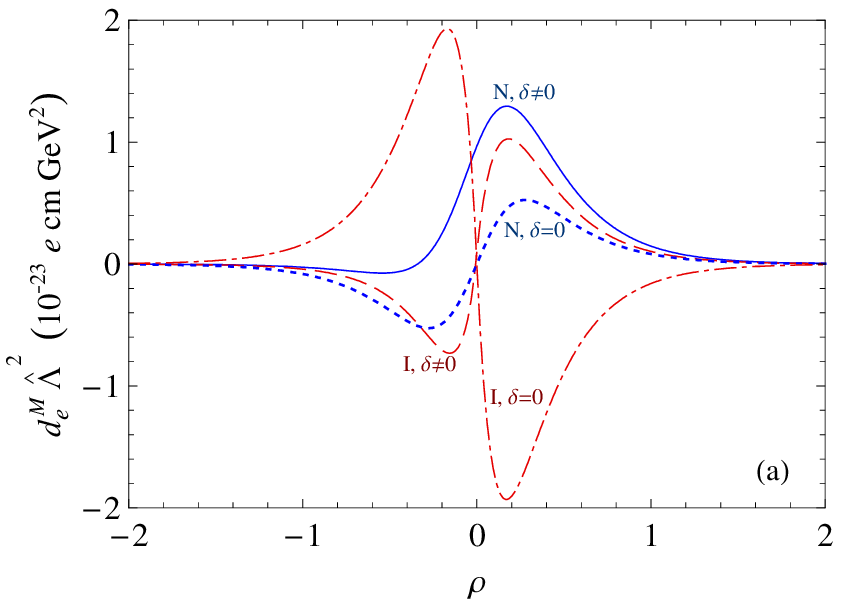} 
\includegraphics[width=87mm]{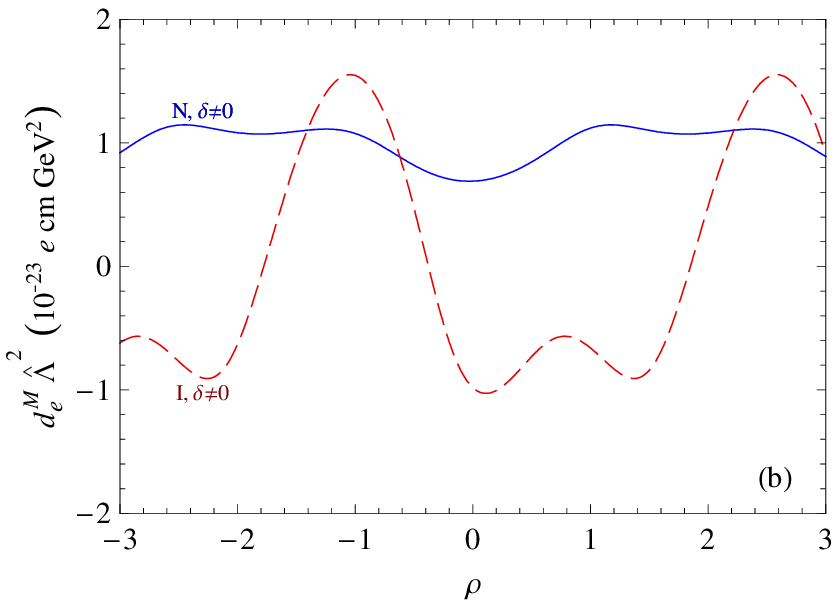} \vspace{-1ex}
\caption{Dependence of $d_e^{\rm M}$ times $\hat\Lambda^2=\Lambda^2/|\xi_{12}^\ell|$\, on $\rho$ 
in the absence of Majorana phases, \,$\alpha_{1,2}=0$,\, for (a)~degenerate $\nu_{i,R}^{}$ and 
complex $O$ and (b)~nondegenerate $\nu_{i,R}^{}$ and real $O$, as explained in the text.
In all figures, the label N (I) refers to the NH (IH) case with \,$m_{1(3)}=0$.\label{de-vs-rho}}  
\end{figure}

For \,$M_\nu={\cal M}\openone$\, and $O$ being complex,
\,${\sf A}=\bigl(2/v^2\bigr){\cal M}_{\,}U_{\scriptscriptstyle\rm PMNS\,}^{}\hat m^{1/2}_\nu
OO^\dagger\hat m^{1/2}_\nu U_{\scriptscriptstyle\rm PMNS}^\dagger$\,
in Eq.\,(\ref{de}).
We can generally write \,$OO^\dagger=e^{2i\sf R}$,\, where $\sf R$ is a~real antisymmetric
matrix with nonzero elements denoted by \,$r_1^{}={\sf R}_{12}=-{\sf R}_{21}$,
\,$r_2^{}={\sf R}_{13}=-{\sf R}_{31}$,\, and \,$r_3^{}={\sf R}_{23}=-{\sf R}_{32}$.
Since $OO^\dagger$ is not diagonal, the Majorana phases in $U_{\scriptscriptstyle\rm PMNS}$
can also enter $\sf A$ if \,$\alpha_{1,2}\neq0$.\,
We focus first on the $CP$-violating effect of $O$ by setting \,$\alpha_{1,2}=0$.\,
For illustration, we pick \,$r_{1,2,3}=\rho$\, and employ the experimental central values
\,$m_2^2-m_1^2=7.54\times10^{-5}{\rm\,eV}^2$,
\,$m_3^2-\frac{1}{2}\bigl(m_1^2+m_2^2\bigr)=2.44_{\,}(2.40)\times10^{-3}{\rm\,eV}^2$,
\,$\sin^2\theta_{12}=0.308$, \,$\sin^2\theta_{23}=0.425_{\,}(0.437)$,
\,$\sin^2\theta_{13}=0.0234_{\,}(0.0239)$,\, and \,$\delta=1.39_{\,}(1.35)\pi$\,
in the NH (IH) case from\,\,\cite{Capozzi:2013csa}.
In Fig.\,\,\ref{de-vs-rho}(a) we present the resulting $d_e^{\rm M}\hat\Lambda^2$ versus
$\rho$ for \,$m_{1(3)}=0$.\,
We also display the curves with \,$\delta=0$\, for comparison.
We have checked that other \,$|r_{1,2,3}|=\rho$\, cases with different relative signs of
$r_{1,2,3}$ produce roughly similar figures and that for \,$m_{1(3)}>0$\, the results tend
to diminish in magnitude.

With \,$\alpha_{1,2}=0$,\, the $CP$-violating effect of $O$ can still occur even if it is
real, provided that $\nu_{i,R}$ are nondegenerate, in which case
\,${\sf A}=\bigl(2/v^2\bigr) U_{\scriptscriptstyle\rm PMNS\,}^{}\hat m^{1/2}_\nu O M_\nu
O^\dagger\hat m^{1/2}_\nu U_{\scriptscriptstyle\rm PMNS}^\dagger$\,
from Eq.\,(\ref{ym}).
For instance, assuming a real $O=e^{\sf R}$ with \,$r_{1,2,3}=\rho$\, and
\,$M_\nu={\cal M}\,{\rm diag}(1,0.8,1.2)$,\, we show the resulting
\,$d_e^{\rm M}\hat\Lambda^2$\, versus $\rho$ in Fig.$\;$\ref{de-vs-rho}(b), where only
the \,$\delta\neq0$\, curves are nonvanishing and  the sinusoidal nature of $d_e$ is visible.
All these examples in Fig.$\;$\ref{de-vs-rho} indicate that $O$ supplies a~potentially 
significant new source of $CP$ violation which can have a bigger impact than $\delta$.

\begin{figure}[b] \vspace{3ex}
\includegraphics[width=87mm]{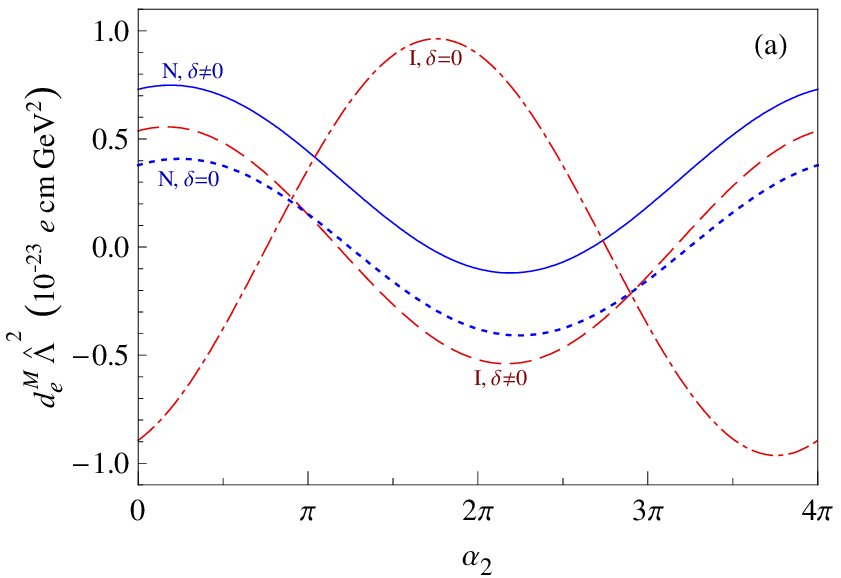} 
\includegraphics[width=87mm]{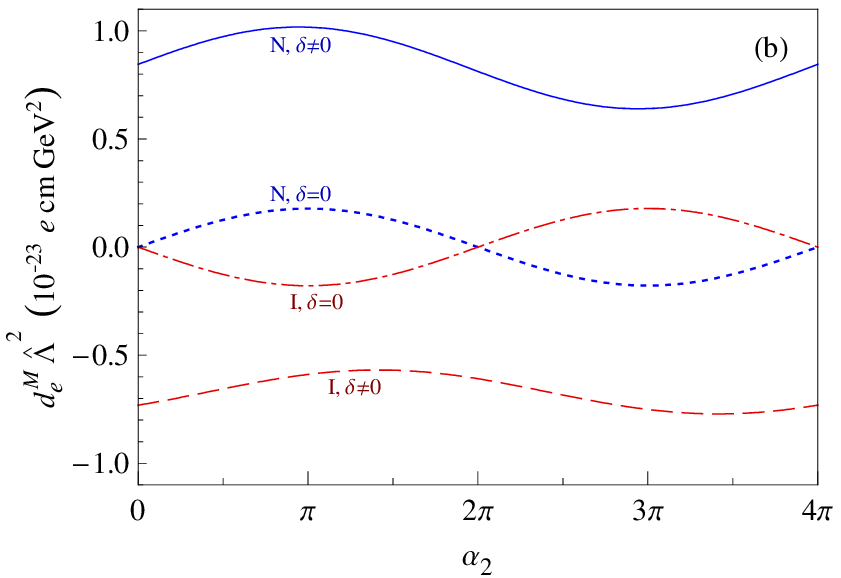} \vspace{-1ex}
\caption{Dependence of \,$d_e^{\rm M}\hat\Lambda^2$\, on $\alpha_2^{}$ for \,$\alpha_1^{}=0$\,
and \,$\rho=0.5$\, with (a)~degenerate $\nu_{i,R}^{}$ and complex $O$ and (b)~nondegenerate
$\nu_{i,R}^{}$ and real $O$, as explained in the text.\label{de-vs-alpha}}
\end{figure}

To see the effect of the Majorana phases, we entertain a couple of possibilities:
(a)~$M_\nu={\cal M}\openone$\, and \,$O=e^{i\sf R}$ and
(b)~$M_\nu={\cal M}\,{\rm diag}(1,0.8,1.2)$\, and \,$O=e^{\sf R}$,\,
both with \,$r_{1,2,3}=\rho$.\,
Fixing \,$\alpha_1^{}=0$,\, we depict the resulting dependence on $\alpha_2^{}$ in
Fig.\,\,\ref{de-vs-alpha} for \,$\rho=0.5$\, and nonzero or zero $\delta$.
Clearly the Majorana phases yield an additional important $CP$-violating contribution to $d_e$
beyond~$\delta$.

Now, the $\xi_{12}^{}$ term contributes not only to $d_e$, but also to the muon anomalous magnetic
moment $(g_\mu-2)$, the radiative decay \,$\mu \to e\gamma$,\, and the \,$\mu\to e$\, conversion.
We have checked that the constraint on $\hat\Lambda$ from the $g_\mu-2$ data is weaker than
that from $|d_e^{\rm exp}|$, but the constraints from the empirical limits on
\,$\mu\to e\gamma$\, and \,$\mu\to e$\, conversion are stronger by up to an order of magnitude
if the $\xi_{12}^{}$ term dominates~\cite{longpaper}.
However, these latter transitions, which are $CP$ conserving, in general also receive
contributions from some of the other $\xi_r^{}$ terms among the operators in Eq.\,(\ref{Leff}),
which can, in principle, reduce the impact of the $\xi_{12}^{}$ term, thereby weakening
the restriction on $\hat\Lambda$ from the data on these processes.
It follows that  $d_e$ provides the best $CP$-violating probe for $\hat\Lambda$.

Lastly,  if $d_e$ is negligible, the ACME experiment~\cite{acme} can probe the $CP$-violating
electron-nucleon interaction~\cite{He:1989xj}
\,${\cal L}_{eN}=-i C_S G_{\rm F\,}\bar e\gamma_5^{}e\,\bar N N/\sqrt2$.\,
At lowest order, it arises mainly from~\cite{Cirigliano:2005ck,longpaper}
\begin{eqnarray}
{\cal L}_{\ell q} \,\,=\,\, \frac{1}{\Lambda^2} \bigl(
\bar E_{R\,}^{}Y_e^\dagger\Delta_{1\,}'L_L^{}\,\bar U_R^{}Y_u^\dagger i\tau_2^{}Q_L^{}
\,+\, \bar E_{R\,}^{}Y_e^\dagger\Delta_{2\,}'L_L^{}\,
\bar Q_L^{}Y_{d\,}^{}D_R^{} \bigr)
\;+\; {\rm H.c.} ~.
\end{eqnarray}
To determine the coupling constant $C_S$, we need the matrix elements
\,$\langle N|m_q^{}\bar q q|N\rangle=g_q^N\bar u_N^{}u_N^{}v$.\,
Employing the chiral Lagrangian estimate in\,\,\cite{He:2008qm},\footnote{These matrix elements
have also been computed with lattice QCD~\cite{Junnarkar:2013ac}, and the results tend to be
smaller than those of chiral Lagrangian calculations and some other
methods~\cite{deAustri:2013saa}.}  we have for \,$M_\nu^{}={\cal M}\openone$\, and real $O$
\begin{eqnarray}
C_S^{} \,=\, \frac{16\sqrt2\,{\cal M}^3m_e^{}}{\Lambda^{2\,}G_{\rm F\,}v^9}
\bigl(m_\mu^2-m_\tau^2\bigr)\bigl(m_1^{}-m_2^{}\bigr)\bigl(m_2^{}-m_3^{}\bigr)
\bigl(m_3^{}-m_1^{}\bigr)\,\bar\xi_{12\,}^\ell J_\ell^{} \,,
\end{eqnarray}
where~\cite{longpaper}
\,$\bar\xi_{12}^\ell=\bigl(g_d^N+g_s^N+g_t^N\bigr)\,\xi_{12}^{\prime(2)}
- \bigl(g_u^N+2g_t^N\bigr)\,\xi_{12}^{\prime(1)}$.\,
Then, with \,${\cal M}\simeq6.2\times10^{14}\,$GeV,\, the maximal values of $g_q^N$
from\,\,\cite{He:2008qm}, and \,$m_{1(3)}=0$\, in the NH (IH) case, the ACME bound
\,$|C_S|<5.9\times10^{-9}$\,~\cite{acme} implies \,$\hat\Lambda'>0.27\,(0.091){\rm\;GeV}$.\,
This is far less stringent than the restriction from $d_e$ directly.

In conclusion, we have studied the electron EDM $d_e$ under the MFV hypothesis.
In this framework, $d_e$ can approach its experimental limit if neutrinos are Majorana
particles acquiring mass via the seesaw mechanism.
Accordingly, $d_e$ is sensitive to the MFV scale of order hundreds of GeV or higher.
We have demonstrated that $d_e$ has the potential to probe not only the Dirac phase in
the lepton mixing matrix, but also the Majorana phases therein and extra $CP$-violation
sources in the Yukawa couplings of the right-handed neutrinos.
Finally, we remark that a~similar analysis on the neutron EDM yields a much weaker
constraint on the MFV scale~\cite{longpaper}.

\acknowledgments

This research was supported in part by the MOE Academic Excellence Program (Grant No. 102R891505)
and NSC of ROC and by NNSF (Grant No. 11175115) and Shanghai Science and Technology Commission
(Grant No. 11DZ2260700) of PRC.

\end{document}